\documentclass[12pt,axodraw]{JHEP3}

\usepackage{axodraw}

\title{Radiative Mechanism to Light Fermion Masses in the MSSM}
\author{C.M. Maekawa and M. C. Rodriguez \\
\textit{Funda\c c\~ao Universidade Federal do Rio Grande-FURG \\
Departamento de F\'\i sica \\
Av. It\'alia, km 8, Campus Carreiros \\
96201-900, Rio Grande, RS \\
Brazil}}

\abstract{
In a previous work we applied a discrete symmetry  (${\cal Z}_{2}^{\prime}$) in order to light fermions acquire mass only at one loop level. This symmetry and the assumption of alignment between fermions and sfermions allow us to avoid FCNC problems.
Here a more general hypothesis of flavor mixing in the sfermion sector of MSSM is considered and we show that the $s$ quark is heavier than $u,d$ quarks due to different content of sfermions contributions. Our results are in agreement with the experimental constraint on the values of sfermions masses.
}

\begin{document}

\section{Introduction}

Family problems of elementary particles have shown to be a challenge since
one realizes that the strong interaction respects isospin, lepton and baryon
numbers conservation. Due to excess of baryons over antibaryons in our
universe the baryon number conservation were pointed to be broken \cite{Sakarov67}. 
Later Weinberg points to lepton and baryon numbers
conservations do not need to be a prior assumption in the framework of Grand
Unified Theories (GUT) where the processes are mediated by superheavy
particles with mass $M\simeq 10^{14}$ GeV \cite{Weinb79}. However the non
observation of proton decay \cite{exp1}, Electric Dipole Moment of
elementary particles \cite{edm} and neutron-antineutron oscillations \cite{israel} 
points to a non trivial violation mechanism of these symmetries.

To this puzzle one can joint questions about the mass generation mechanism
which are able to describe the observed mass hierarchy of particles and
mixing angles. A known mechanism is based on Yukawa couplings between fermions
and scalars of the models (Standard Model (SM), Supersymmetric models (SUSY)
and GUT) but among the possibilities of such couplings there are sources for
dangerous Flavor Change Neutral Currents (FCNC) processes, like proton decay and
neutron-antineutron oscillation.

Besides, the recent data from neutrinos experiments add more questions: What
is their mass scale? Which is their mass organization pattern?

The SM describes the family structure as doubletes of $SU\left( 2\right) $ 
\cite{SM} and it has been able to described most of present day data. In the
case of mass generation mechanism, SM tells us that fermions obtain their
masses through Yukawa couplings with Higgs doublet while neutrinos have no
mass. However the values of these couplings constants remain arbitrary.\
There are also other aspects which cannot be explain in the framework of SM,
e.g., the non-leptonic without strangeness changing processes. In this case
the problem is not due to family structure, it is the interplay between
strong and week interaction. The strong repulsive core keeps the nucleons
away from each other at a distance enough to prevent the gauge bosons
exchanges between quarks \cite{Adelb+85}. In this case one has to deal with
nucleons and pions degrees of freedom instead of quarks and gluons. A
reliable and consistent description with underline principles is obtained in
the framework of Chiral Perturbation Theory (ChPT) \cite{SMChPT}.

In order to determine the values of Yukawa coupling constants or at least to find a
way to constrain them there are approaches based on GUT +SUSY +$G_{f}$ \cite{ReviewBerez} 
or SM +$G_{f}$ \cite{Moha+2006} where $G_{f}$ is an additional
family symmetry required in order to constrain these constants. On GUT
+SUSY +$G_{f}$ approaches, the masses of fermions are degenerate at GUT
scale and a mass generation mechanism based on renormalization group
equations gives rise to the hierarchy pattern observed at Fermi scale. One
classifies quarks as $u$ type ($t$, $c$, $u$) and $d$ type ($b$, $s$, $d$)
and the hierarchy pattern follows a power law, e.g.: for horizontal
hierarchy:
\begin{equation}
\begin{tabular}{lll}
$m_{t}:m_{c}:m_{u}\sim 1:\varepsilon _{u}:\varepsilon _{u}^{2}$ &  & $\varepsilon _{u}\simeq 500^{-1}$ \\ 
$m_{b}:m_{s}:m_{d}\sim 1:\varepsilon _{d}:\varepsilon _{d}^{2}$ &  & $\varepsilon _{d}\simeq 50^{-1}$ \\ 
$m_{\tau }:m_{\mu }:m_{\varepsilon }\sim 1:\varepsilon _{e}:\varepsilon_{e}^{2}$ &  & $\varepsilon _{e}\simeq 50^{-1}$
\end{tabular}
\end{equation}
where $m_{u}$ and $m_{d}$ are the current quark mass.

Another source of flavor problems is the misalignment of fermion- sfermion
couplings. It is due to the transformation that diagonalizes the fermion
mass matrix does not simultaneously diagonalize the corresponding sfermion
mass squared matrices. The lack of observation of the decays $\mu
\rightarrow e\gamma $, $\tau \rightarrow \mu \gamma $ and $\tau \rightarrow
e\gamma $ put some constraints on the lepton-slepton coupling. On the other
hand, processes like $b\rightarrow s\gamma $ decay and the measurements of
mass difference in $B^{0}\bar{B}^{0}$ and $D^{0}\bar{D}^{0}$ yield
constraints on the quark-squark couplings, the most stringent restrictions
here come from one knows about $K^{0}\bar{K}^{0}$ mixing.

In general there are three ways to suppress this problem \cite{dress,tata}:

\begin{enumerate}
\item[a] Arrange for degeneracy or universality of masses of sfermions with
the same quantum numbers. In this scenario the $K^{0}\bar{K}^{0}$ mixing
expression is suppressed because the $\Delta m_{\tilde{d}_{\imath}}$ is very
small;

\item[b] One can assume an alignment between the fermion and sfermion mass
matrices so that both can be made diagonal in the same basis. In this case,
the fermion and sfermion mass matrices is said to be aligned. Such an
alignment is included in models with so-called horizontal symmetries which
links the various generations;

\item[c] The third choice is to take the masses of sfermions of the first
two generations to be very large, in the multi-TeV range. This solution to
the SUSY flavor problem is known as decoupling.
\end{enumerate}

One may also consider various combinations of these options. The flavor
violating contributions have been parametrized by Gabbiani \textit{et al} 
\cite{9604378} in the framework of mass insertion method. In this approach
one works in a basis where the mass matrix of fermions of a given charge as
well as the corresponding fermion-sfermion-neutral gaugino couplings are
diagonal in flavor space. Flavor violation is then described by flavor
non-diagonal entries and the constraints are expressed as bounds on the
dimensionless quantities.

The first attempt to apply the radiative mechanism of mass generation to
the light fermions was suggested by S. Weinberg \cite{Weinberg:1971nd,Weinberg:1972ws}. 
Later L.~Iba\~{n}ez shows if SUSY is
spontaneously broken one generates only tiny small fermion masses
radiatively \cite{Ibanez:1982xg}. In order to restrict this mechanism to
the first family a discrete symmetry is applied into SUSY models in Refs. 
\cite{banks,ma}. From the analysis performed by Ferrandis \cite{ferrandis1,ferrandis2} 
the radiative mechanism of mass generation of
fermions is allowed through sfermion--gaugino loops and the observed flavor
physics is obtained if \textquotedblleft the supersymmetry breaking terms
receive small corrections, which violate the symmetry of the superpotential".

In a previous work \cite{cmmc} we followed a mass pattern of Chiral
descriptions \cite{SMChPT} where a Chiral scale ($\Lambda _{\chi }=1$ GeV)
allow us to classify the quarks as light ones ( $u,d$ and $s$) and heavy
ones ($c$, $t$ and $b$). Thus we introduced a $\mathcal{Z}_{2}^{\prime }$
symmetry in the MSSM and in the SUSY Left-Right Models in order to allow the
light quarks acquire mass only by means of radiative mass generation
mechanism \cite{banks,ma,Pok+90} while the FCNC problems are avoided . The
Chiral mass hierarchy pattern and a consistent picture with experimental
data of Cabibbo-Kobayashi-Maskawa (CKM) matrix were obtained. We also
showed that under $\mathcal{Z}_{2}^{\prime }$ symmetry, a similar pattern for
electron, muon and tau can be obtained. The heavy leptons ($\mu $ and $\tau $) 
acquire mass at tree level while the electron acquires mass at 1-loop
level . We also assumed the alignment of squarks with the quarks and due to
the absence of inter family mixing of squarks each quark receives
contribution only from its corresponding supersymmetric partner and in order
to describe the mass gap between strange and non strange quarks we need to
consider the strange supersymmetric partner heavier than non-strange
supersymmetric partners.

In this work we remove the assumption of alignment between quark and squarks
sectors and we explore the effects of $\mathcal{Z}_{2}^{\prime }$ symmetry 
\footnote{We review this symmetry in section \ref{sec:ma}} on the masses of sfermions
in section \ref{SpartMassReview}. In the section \ref{sec:soft} the mass of
light fermions are re-evaluated and the contribution of sfermions are still
different to each mass of light fermions.

We present at section \ref{sec:numerical} the masses of the light fermions.
From these results, we can explain why the quarks $u$ and $d$ are lighter
than the $s$ quark. Our notation is shown in the appendix \ref{append:mssm}.
The details of computations are presented in our appendices (\ref{apend:fsg}
- \ref{append:feynmanaintegration}).

\section{$\mathcal{Z}_{2}^{\prime}$ symmetry in the MSSM}

\label{sec:ma}

In our previous work, we introduced the following $\mathcal{Z}_{2}^{\prime }$
symmetry on the Lagrangian of the MSSM \cite{cmmc} 
\begin{equation}
\widehat{d}_{2L}^{c}\longrightarrow -\widehat{d}_{2L}^{c},\,\ 
\widehat{d}_{3L}^{c}\longrightarrow -\widehat{d}_{3L}^{c},\,\ 
\widehat{u}_{3L}^{c}\longrightarrow -\widehat{u}_{3L}^{c},\,\ 
\widehat{l}_{3L}^{c}\longrightarrow -\widehat{l}_{3L}^{c},  
\label{z2def}
\end{equation}
while the others superfields of the model \footnote{Our notation and the particles contents of this model are 
shown in Appendix \ref{append:mssm}.} are even under this symmetry.

The invariant superpotential under $\mathcal{Z}_{2}^{\prime }$ and $R$-parity symmetries is given by 
\begin{eqnarray}
W_{\mathrm{R-inv}}^{\mathcal{Z}_{2}^{\prime }\mathrm{even}} &=&\mu 
\hat{H}_{1}\hat{H}_{2}+\sum_{\imath =1}^{3}y_{\imath 1}^{d}\hat{Q}_{\imath L}\hat{H}_{1}\hat{d}_{1L}^{c}
+\sum_{\imath =1}^{3}\sum_{\jmath =1}^{2}y_{\imath \jmath}^{l}\hat{L}_{\imath L}\hat{H}_{1}\hat{l}_{\jmath L}^{c}
+\sum_{\imath =1}^{3}\sum_{\jmath =1}^{2}y_{\imath \jmath}^{u}\hat{Q}_{\imath L}\hat{H}_{2}\hat{u}_{\jmath L}^{c}.  
\nonumber  \label{allowz2} \\
&&
\end{eqnarray}
The $R$-parity symmetric but $\mathcal{Z}_{2}^{\prime }$ forbidden terms are
given by: 
\begin{equation}
W_{\mathrm{R-inv}}^{\mathcal{Z}_{2}^{\prime }\mathrm{odd}}=\sum_{\imath =1}^{3}
\sum_{\jmath =2}^{3}y_{\imath \jmath}^{d}\hat{Q}_{\imath L}\hat{H}_{1}\hat{d}_{\jmath L}^{c}+
\sum_{\imath =1}^{3}y_{\imath 3}^{l}\hat{L}_{\imath L}\hat{H}_{1}\hat{l}_{3L}^{c}+
\sum_{\imath =1}^{3}y_{\imath 3}^{u}\hat{Q}_{\imath L}\hat{H}_{2}\hat{u}_{3L}^{c}.
\label{forbidz2}
\end{equation}

As a consequence of Eq.(\ref{allowz2}), the fermions $u,d,s$ and $e$
are prevented to acquire mass at tree level in a way similar as presented in
Refs.\cite{banks,ma,ferrandis1,ferrandis2}. These fermions are massless
due to the absence of the terms showed in Eq.(\ref{forbidz2}). On the other hand,
Supersymmetric non-renormalization theorem guarantee that corrections to the
fermions masses are very small, even if the
discrete symmetry (\ref{z2def}) is broken.

An interesting question we don't deal in our first work is: How Does our $\mathcal{Z}_{2}^{\prime }$ 
symmetry act on the sfermion sector? The answer can be obtained from  Eq.(\ref{z2def}). The
behavior of scalar components of chiral superfields under $\mathcal{Z}_{2}^{\prime }$ 
symmetry are given by: 
\begin{eqnarray}
\tilde{d}_{2}^{c} &\longrightarrow &-\tilde{d}_{2}^{c},\,\
d_{2}^{c}\longrightarrow -d_{2}^{c},\,\ 
\tilde{d}_{3}^{c}\longrightarrow - \tilde{d}_{3}^{c},\,\ d_{3}^{c}\longrightarrow -d_{3}^{c},  \nonumber \\
\tilde{u}_{3}^{c} &\longrightarrow &-\tilde{u}_{3}^{c},\,\
u_{3}^{c}\longrightarrow -u_{3}^{c},\,\ 
\tilde{l}_{3}^{c}\longrightarrow -\tilde{l}_{3}^{c},\,\ l_{3}^{c}\longrightarrow -l_{3}^{c},  
\label{z2comp}
\end{eqnarray}
while all other fields of the model are even. It worth noting that the $\mathcal{Z}_{2}^{\prime }$ 
symmetry has the same role as in the fermion
sector: it forbids the flavor mixing between the third family and the other
two families of sfermions.

As we show below, we also obtain the following features: because the couplings
between the squarks from the third family with the other two families are
forbidden, the assumption of alignment between quark and squark sector
can be removed. Then a particular texture for mass matrix of squarks
consistent with physical bounds comes out. Therefore the $\mathcal{Z}_{2}^{\prime }$ 
symmetry help us to keep under control the dangerous FCNC
problems and we still obtain the mass hierarchy pattern without any additional
assumptions.

Another feature in the $\mathcal{Z}_{2}^{\prime }$ symmetric case is the null value for 
EDM of electron and of neutron. Because of the symmetry the 
left-right mixing angle vanished in the sleptons and squarks sectors. 
These mixing angles contribute to the EDM calculation and in
this case there are no contributions to the EDM coming from the MSSM.
Therefore, the only contribution to the EDM of these particles come from SM.

\section{Masses of the supersymmetric Particles}

\label{SpartMassReview}

The discussion in this section is based on review articles of Refs.\cite{dress,tata,mssm,balin,mussolf}. 
We start with a general study of mass
generation of supersymmetric particles. The reason to perform this study is
due to the fact that masses and mixing of sparticles are of crucial
importance both experimentally and theoretically: i) they determine the
properties of the sparticles searched for and ii) they are directly related
to the question of how SUSY is broken \cite{dress,tata}.

Once $SU(2)_{L}\otimes U(1)_{Y}$ symmetry is broken, fields with the same 
$SU(3)_{c}\otimes U(1)_{em}$ quantum numbers (and, of course, R-parity, 
$\mathcal{Z}_{2}^{\prime }$ and spin) can mix with each other. In the
framework of Standard Model, $B^{0}$ and $W^{i}$ mix to $\gamma $, $Z^{0}$,
and $W^{\pm }$ is an example of this kind of mixing. Also the Dirac masses
of quarks and leptons can be understood as such mixing terms. For the case
of MSSM, this mixing also affects squarks, sleptons, Higgs bosons, as well
as gauginos and higgsinos. The only exception is the gluino, which is the
only color octet fermion in the model.

\subsection{Super-CKM basis for Sfermions}

\label{apend:sfermions}

There is no longer alignment assumption and we perform the diagonalization
procedure in the Super-CKM (SCKM) basis. Here we present the relevant
equations for our work and a detailed discussion can be found at Ref.~\cite{Chung:2003fi}.

Likewise we have done in our superpotential 
\footnote{See Eqs.(\ref{allowz2},\ref{forbidz2})}, we separate the soft SUSY breaking
terms into two terms: 
\begin{equation}
\mathcal{L}_{\mathrm{soft}}=
\mathcal{L}_{\mathrm{soft}}^{\mathcal{Z}_{2}^{\prime }\mathrm{even}}+
\mathcal{L}_{\mathrm{soft}}^{\mathcal{Z}_{2}^{\prime }\mathrm{odd}}
\end{equation}
where $\mathcal{L}_{\mathrm{soft}}^{\mathcal{Z}_{2}^{\prime }\mathrm{even}}$
is the even component under $\mathcal{Z}_{2}^{\prime }$ (Eq.(\ref{z2comp}))
and it reads: 
\begin{eqnarray}
\mathcal{L}_{\mathrm{soft}}^{\mathcal{Z}_{2}^{\prime }\mathrm{even}} &=&-
\frac{1}{2}\left( \sum_{\imath =1}^{8}m_{\tilde{g}}\lambda _{C}^{\imath}\lambda_{C}^{\imath}+
\sum_{p=1}^{3}m_{\lambda }\lambda _{A}^{p}\lambda_{A}^{p}+m^{\prime }\lambda _{B}\lambda _{B}
+h.c.\right) -\left[
\sum_{\imath =1}^{3}\sum_{\jmath =1}^{3}\tilde{L}_{\imath L}^{\star }\left( M_{L}^{2}
\right)_{\imath \jmath}\tilde{L}_{\jmath L}\right.  \nonumber \\
&+&\left. \sum_{\imath =1}^{2}\sum_{\jmath =1}^{2}\tilde{l}_{\imath L}^{c\star }\left(
M_{l}^{2}\right) _{\imath \jmath}\tilde{l}_{\jmath L}^{c}+\tilde{l}_{3L}^{c\star }\left(
M_{l}^{2}\right) _{33}\tilde{l}_{3L}^{c}+\sum_{\imath =1}^{3}\sum_{\jmath =1}^{3}
\tilde{Q}_{\imath L}^{\star }\left( M_{Q}^{2}\right) _{\imath \jmath}\tilde{Q}_{\jmath L}\right.  \nonumber
\\
&+&\left. \sum_{\imath =1}^{2}\sum_{\jmath =1}^{2}\tilde{u}_{\imath L}^{c\star }\left(
M_{u}^{2}\right) _{\imath \jmath}\tilde{u}_{\jmath L}^{c}+\tilde{u}_{3L}^{c\star }\left(
M_{u}^{2}\right) _{33}\tilde{u}_{3L}^{c}+\tilde{d}_{1L}^{c\star }\left(
M_{d}^{2}\right) _{11}\tilde{d}_{1L}^{c}+\sum_{\imath =2}^{3}\sum_{\jmath =2}^{3}
\tilde{d}_{\imath L}^{c\star }\left( M_{d}^{2}\right) _{\imath \jmath}\tilde{d}_{\jmath L}^{c}\right] 
\nonumber \\
&-&M_{1}^{2}\tilde{H}_{1}^{\star }\tilde{H_{1}}-
M_{2}^{2}\tilde{H}_{2}^{\star }\tilde{H_{2}}-M_{12}^{2}\left( H_{1}H_{2}+h.c.\right) -\left[
\sum_{\imath =1}^{3}\sum_{\jmath =1}^{2}H_{1}\tilde{L}_{\imath L}
\left( A^{l}\right) _{\imath \jmath}\tilde{l}_{\jmath L}^{c}\right.  \nonumber \\
&+&\left. \sum_{\imath =1}^{3}\sum_{\jmath =1}^{2}H_{2}\tilde{Q}_{\imath L}\left( 
A^{u}\right)_{\imath \jmath}\tilde{u}_{\jmath L}^{c}
+\sum_{\imath =1}^{3}H_{1}\tilde{Q}_{\imath L}\left( A^{d}\right)_{\imath 1}\tilde{d}_{1L}^{c}+h.c.\right] .  \label{lsoftpar}
\end{eqnarray}
The $m^{\prime }$, $m_{\lambda }$, and $m_{\tilde{g}}$ are U(1), SU(2) and
SU(3) gaugino masses respectively. The mass terms of Higgs fields are
denoted by $M_{1}^{2}$, $M_{2}^{2}$, and $M_{12}^{2}$. The symbol ($^{\star} $) 
in a scalar field is the charge conjugate of this field, it means we
take their anti-particle.

The components whose also break $\mathcal{Z}_{2}^{\prime }$ symmetry 
($\mathcal{L}_{\mathrm{soft}}^{\mathcal{Z}_{2}^{\prime }\mathrm{odd}}$) are
given by 
\begin{eqnarray}  \label{rgggg}
\mathcal{L}_{\mathrm{soft}}^{\mathcal{Z}_{2}^{\prime }\mathrm{odd}} &=&-
\left[ \sum_{\imath =1}^{2}\tilde{l}_{\imath L}^{c\star }\left( M_{l}^{2}\right) _{\imath 3}
\tilde{l^{c}}_{3L}+\sum_{\imath =1}^{2}\tilde{u}_{\imath L}^{c\star }\left(
M_{u}^{2}\right) _{\imath 3}\tilde{u^{c}}_{3L}+\sum_{\imath =2}^{3}
\tilde{d}_{\imath L}^{c\star }\left( M_{d}^{2}\right) _{\imath 1}\tilde{d^{c}}_{1L}\right] 
\nonumber  \label{lsoftimpar} \\
&-&\left[ \sum_{\imath =1}^{3}\sum_{\jmath =2}^{3}A_{\imath \jmath}^{d}H_{1}\tilde{Q}_{\imath L}
\tilde{d}_{\jmath L}^{c}+\sum_{\imath =1}^{3}A_{\imath 3}^{u}H_{2}\tilde{Q}_{\imath L}
\tilde{u}_{3L}^{c}+\sum_{\imath =1}^{3}A_{\imath 3}^{l}H_{1}\tilde{L}_{\imath L}\tilde{l}_{3L}^{c}+h.c.
\right] .  \nonumber \\
&&
\end{eqnarray}
It worth remembering the scalar masses $M_{Q}^{2}$, $M_{u}^{2}$, $M_{d}^{2}$
, $M_{L}^{2}$, and $M_{l}^{2}$ are in general $3\!\times \!3$ hermitian
matrices in generation space, while $A^{u}$, $A^{d}$, and $A^{l}$ are
general $3\!\times \!3$ matrices. Allowing all the parameters in 
Eqs.(\ref{lsoftpar},\ref{lsoftimpar}) to be complex, we end up with 124 masses,
phases and mixing angles as free parameters of the model.

Constraints from FCNC processes also restrict the form of the soft SUSY
breaking trilinear terms $A^{u},A^{d}$ and $A^{e}$. For example, the data from 
$K^{0}\bar{K}^{0}$ mixing require the off-diagonal entries of the $A^{d}$
matrix to be small.

Besides, these terms make contributions to fermion masses \cite{dress,tata,banks,ma,ferrandis1,ferrandis2,cmmc}. 
The requirement that
contributions to the fermion masses to be smaller than the fermion masses
themselves put tight constraints to the magnitudes of the $A$-terms.

As we said at section ~\ref{sec:ma} if our $\mathcal{Z}_{2}^{\prime }$ is
hold there is no contribution to the EDM. However we have to break this
symmetry in order to generate masses to the fermions and we also allow
contributions to the EDM. Limits on the imaginary part of the soft SUSY
breaking $A$-terms can be obtained from experimental upper limits of
electron and neutron EDM \cite{tata} 
\begin{eqnarray}
d_{e} &\propto &\sqrt{\Im (A^{e})v_{1}}<6\cdot 10^{-4}m_{\tilde{e}}, 
\nonumber \\
d_{n} &\propto &\sqrt{\Im (A^{d})v_{1}}<0,002m_{\tilde{d}}.
\end{eqnarray}

Here, we consider the most general scenario which is called in the
literature as non-minimal flavor scenario. We follow reference \cite{Chung:2003fi} 
and the sfermions fields are rearranged into the following
vector with six component: 
\begin{equation}
\tilde{f}^{T}=\left( \tilde{f}_{\imath L}\,\ \tilde{f}_{\imath R}\right) \,\ ,
\label{squarkvector}
\end{equation}
where each $\tilde{f}_{\imath L},\tilde{f}_{\imath R}$ is a three component column
vector in generation space \footnote{We want to emphasize that $\tilde{f}$ is the superpartner of any matter
fermion field $f$.}, $f=u,d,l$ and $\imath =1,2,3$. Then we can write sfermion
mass term of the MSSM Lagrangian in the following way: 
\begin{equation}
\tilde{f}^{\dagger }\mathcal{M}_{\tilde{f}}^{2}\tilde{f},
\label{squarkmassmatrix}
\end{equation}
where $\mathcal{M}_{\tilde{f}}^{2}$ are $6\!\times \!6$ sfermion mass
matrices --- one for up--type, one for down--type squarks and one for
charged sleptons.

The general squared mass matrix of sfermions can be written as a $2\times 2$
Hermitian matrix of $3\times 3$ blocks in the space spanned by the vector of
Eq.(\ref{squarkvector}) \cite{dress} 
\begin{equation}
\mathcal{M}_{\tilde{f}}^{2}=\left( 
\begin{array}{cc}
\mathcal{M}_{\tilde{f}_{LL}}^{2} & \mathcal{M}_{\tilde{f}_{LR}}^{2} \\ 
\mathcal{M}_{\tilde{f}_{LR}}^{2\dagger } & \mathcal{M}_{\tilde{f}_{RR}}^{2}
\end{array}
\right) .  \label{sfermion3}
\end{equation}

The squared mass matrix of sfermions are diagonalized by pairs of $3 \times
6 $ matrices as follows: 
\begin{eqnarray}
\mathrm{diag}(m^2_{\widetilde{u}_1} \ldots m^2_{\widetilde{u}_6})= \left ( 
\begin{array}{cc}
W^{\tilde{u}_{L}\dagger} & W^{\tilde{u}_{R} \dagger}
\end{array}
\right ) \mathcal{M}^{2}_{\tilde{u}} \left ( 
\begin{array}{c}
W^{\tilde{u}_{L}} \\ 
W^{\tilde{u}_{R}}
\end{array}
\right ),
\end{eqnarray}
\begin{eqnarray}
\mathrm{diag}(m^2_{\widetilde{d}_1} \ldots m^2_{\widetilde{d}_6})= \left ( 
\begin{array}{cc}
W^{\tilde{d}_{L}\dagger} & W^{\tilde{d}_{R} \dagger}
\end{array}
\right ) \mathcal{M}^{2}_{\tilde{d}} \left ( 
\begin{array}{c}
W^{\tilde{d}_{L}} \\ 
W^{\tilde{d}_{R}}
\end{array}
\right ),
\end{eqnarray}
\begin{eqnarray}
\mathrm{diag}(m^2_{\widetilde{l}_1} \ldots m^2_{\widetilde{l}_6})= \left ( 
\begin{array}{cc}
W^{\tilde{E}_{L}\dagger} & W^{\tilde{E}_{R} \dagger}
\end{array}
\right ) \mathcal{M}^{2}_{\tilde{l}} \left ( 
\begin{array}{c}
W^{\tilde{E}_{L}} \\ 
W^{\tilde{E}_{R}}
\end{array}
\right ).
\end{eqnarray}

However, it is common to rotate quarks to their mass eigenstates basis and
to rotate squarks in exactly the same way as quarks. This is the so-called
Super-CKM (SCKM) basis. It is a suitable basis for the study of flavor
violation process since all the unphysical parameters in the Yukawa matrices
have already been rotated away, see at Ref.~\cite{Chung:2003fi}.

\subsubsection{The Squarks}

\label{apend:squarks}

Here we present the constraints on the elements of squarks mass matrix due
to our $\mathcal{Z}_{2}^{\prime}$ symmetry. It is worth recalling the
hypothesis of misalignment between the squark and quark mass matrices is
present in the most general parameterization of the MSSM and it generates
dangerous FCNC effects in conflict with experiment. Specially, the data on 
$K^{0}-\bar{K}^{0}$ and $D^{0}-\bar{D}^{0}$ mixing impose severe constraints
on the mixing involving the $u$-squark and $d$-squark~\cite{9604378}.

However, as discussed at the beginning of this section, only particles with
the same quantum number can mix with each other. On the other hand, $H_{1,2}$
are even under the $\mathcal{Z}_{2}^{\prime }$ symmetry. Thus 
$\mathcal{Z}_{2}^{\prime }$ symmetry is maintained in the presence of the spontaneous
breaking of gauge symmetry as can be shown by Eq.(\ref{lsoftpar}).

It is useful to stress the parameters $A_{ib}^{d}$ and $A_{i3}^{u}$ (Eq.(\ref{rgggg})) 
should be zero because they are not allowed by our 
$\mathcal{Z}_{2}^{\prime }$ symmetry. This means the third family does not mix with
other two families of squarks, then the following matrix elements of Eq.(\ref
{sfermion3}) vanished: 
\begin{eqnarray}
(\mathcal{M}_{\tilde{f}_{LR}}^{2})_{31} &=&
(\mathcal{M}_{\tilde{f}_{LR}}^{2})_{32}=(\mathcal{M}_{\tilde{f}_{LR}}^{2})_{34}=
(\mathcal{M}_{\tilde{f}_{LR}}^{2})_{35}=0, \\
(\mathcal{M}_{\tilde{f}_{LR}}^{2})_{61} &=&
(\mathcal{M}_{\tilde{f}_{LR}}^{2})_{62}=(\mathcal{M}_{\tilde{f}_{LR}}^{2})_{64}=
(\mathcal{M}_{\tilde{f}_{LR}}^{2})_{65}=0,  \nonumber
\end{eqnarray}
and we obtain the same texture of mass matrix of squark as in reference \cite{mariana5}, 
but only with the requirement of invariance under $\mathcal{Z}_{2}^{\prime }$: 
\begin{equation}
\!\!\!\!\!\!\!\!\mathcal{M}_{\tilde{u}\{\tilde{d}\}}^{2}=\left( 
\begin{array}{cccccc}
M_{{\tilde{L}}c\{s\}}^{2} & (M_{\tilde{U}\{\tilde{D}\}}^{2})_{LL} & 0 & 
m_{c\{s\}}\mathcal{A}_{c\{s\}} & (M_{\tilde{U}\{\tilde{D}\}}^{2})_{LR} & 0
\\ 
(M_{\tilde{U}\{\tilde{D}\}}^{2})_{LL} & M_{{\tilde{L}}t\{b\}}^{2} & 0 & 
(M_{\tilde{U}\{\tilde{D}\}}^{2})_{RL} & m_{t\{b\}}\mathcal{A}_{t\{b\}} & 0 \\ 
0 & 0 & M_{{\tilde{L}}u\{d\}}^{2} & 0 & 0 & m_{u\{d\}}\mathcal{A}_{u\{d\}} \\
[0.3ex] 
m_{c\{s\}}\mathcal{A}_{c\{s\}} & (M_{\tilde{U}\{\tilde{D}\}}^{2})_{RL} & 0 & 
M_{{\tilde{R}}c\{s\}}^{2} & (M_{\tilde{U}\{\tilde{D}\}}^{2})_{RR} & 0 \\ 
(M_{\tilde{U}\{\tilde{D}\}}^{2})_{LR} & m_{t\{b\}}\mathcal{A}_{t\{b\}} & 0 & 
(M_{\tilde{U}\{\tilde{D}\}}^{2})_{RR} & M_{{\tilde{R}}t\{b\}}^{2} & 0 \\ 
0 & 0 & m_{u\{d\}}\mathcal{A}_{u,\{d\}} & 0 & 0 & M_{{\tilde{R}}u\{d\}}^{2}
\end{array}
\right) ,  \label{eq:squarkmassmariana}
\end{equation}
with 
\begin{eqnarray}
M_{{\tilde{L}}q}^{2} &=&M_{Q,q}^{2}+m_{q}^{2}+\cos 2\beta
(T_{q}-Q_{q}s_{W}^{2})M_{Z}^{2}\,,  \nonumber  \label{eq:squarkparam} \\
M_{{\tilde{R}}\{u,c,t\}}^{2} &=&M_{u,\{u,c,t\}}^{2}+m_{u,c,t}^{2}+\cos
2\beta Q_{t}s_{W}^{2}M_{Z}^{2}\,,  \nonumber \\
M_{{\tilde{R}}\{d,s,b\}}^{2} &=&M_{d,\{d,s,b\}}^{2}+m_{d,s,b}^{2}+\cos
2\beta Q_{b}s_{W}^{2}M_{Z}^{2}\,, \\
\mathcal{A}_{u,c,t} &=&\sum_{\imath =1}^{3}A_{3\imath ,2\imath ,1\imath}^{u,c,t}-\mu \cot \beta
\,,\;\;\;\;\mathcal{A}_{d,s,b}=\sum_{\imath =1}^{3}A_{3\imath ,2\imath ,1\imath}^{d,s,b}-\mu \tan
\beta \,,  \nonumber
\end{eqnarray}
where $m_{q}$, $T_{q}$, $Q_{q}$ are, respectively, the mass, isospin, and
electric charge of the quark $q$, $M_{Z}$ is the mass of $Z$-boson, 
$s_{W}\equiv \sin \theta _{W}$ and $\theta _{W}$ is the electroweak mixing
angle. The masses $m_{u}$ and $m_{d}$ are null, we keep them here only to
show that in diagonalization procedure they give rise to the mixing in the
third family.

The flavor-changing entries are contained in 
\begin{eqnarray}  \label{pok1a}
\begin{array}{ccc}
(M^2_{\widetilde{U}})_{LL} = V_{U_L} M^{2^*}_Q V^{\dagger}_{U_L}, 
\hspace{0.5cm} & (M^2_{\widetilde{U}})_{RR} = V_{U_R} M^{2^*}_u V^{\dagger}_{U_R}, 
\hspace{0.5cm} & (M^2_{\widetilde{U}})_{LR} = v_u^*V_{U_L} A_u^{*}
V_{U_R}^{\dagger}, \vspace{0.2cm} \\ 
(M^2_{\widetilde{D}})_{LL} = V_{D_L} M^{2^*}_Q V_{D_L}^{\dagger}, 
\hspace{0.5cm} & (M^2_{\widetilde{D}})_{RR} = V_{D_R} M^{2^*}_d V_{D_R}^{\dagger}, 
\hspace{0.5cm} & (M^2_{\widetilde{D}})_{LR} = v_d^*V_{D_L} A_{d}^{*}
V_{D_R}^{\dagger}.
\end{array}
\nonumber \\
\end{eqnarray}
Eq.(\ref{pok1a}) demonstrates the needs of all four matrices 
$V_{{U,D}_{L,R}} $ even though the observed CKM matrix only constraints one
combination of them.

Each one of general six by six mass matrix of Eq.(\ref{eq:squarkmassmariana}) 
can be split into two matrices: one of order four and the other of order
two. The order four matrix corresponds to the mass and mixing of squarks of
first and second families while the masses and mixing of squarks of third
family constitute the mass matrix of order two. One performs the
diagonalization procedure of the matrices in the following way:

\begin{enumerate}
\item[a-)] {Mixing between the First and Second Family of the Squarks}

The four component vectors for up-squark and down-squarks are, respectively,
($\tilde{u}_{1 L}$,$\tilde{u}_{2L}$, $\tilde{u}_{1R}$,$\tilde{u}_{2R}$) and 
($\tilde{d}_{1L}$,$\tilde{d}_{2L}$ $\tilde{d}_{1R}$,$\tilde{d}_{2R}$). Thus
the squark squared mass matrices are given by: 
\begin{equation}
\mathcal{M}_{\tilde{u}\{\tilde{d}\}}^{2}=\left( 
\begin{array}{llll}
M_{\tilde{L},c\{s\}}^{2} & (M_{\tilde{U}\{\tilde{D}\}}^{2})_{LL} & 
m_{c\{s\}} \mathcal{A}_{c\{s\}} & (M_{\tilde{U}\{\tilde{D}\}}^{2})_{LR} \\ 
(M_{\tilde{U}\{\tilde{D}\}}^{2})_{LL} & M_{{\tilde{L}}t\{b\}}^{2} & 
(M_{\tilde{U}\{\tilde{D}\}}^{2})_{RL} & m_{t\{b\}}\mathcal{A}_{t\{b\}} \\ 
(M_{\tilde{U}\{\tilde{D}\}}^{2})_{LR} & (M_{\tilde{U}\{\tilde{D}\}}^{2})_{RL}
& M_{{\tilde{R}}c\{s\}}^{2} & (M_{\tilde{U}\{\tilde{D}\}}^{2})_{RR} \\ 
(M_{\tilde{U}\{\tilde{D}\}}^{2})_{LR} & m_{t\{b\}}\mathcal{A}_{t\{b\}} & 
(M_{\tilde{U}\{\tilde{D}\}}^{2})_{RR} & M_{{\tilde{R}}t\{b\}}^{2}
\end{array}
\right) .  \label{eq.usquarkmass}
\end{equation}

In order to diagonalize $\mathcal{M}_{\tilde{u}\{\tilde{d}\}}^{2}$ one
requires two rotation $4\times 4$ matrices: one for the $up$-squarks 
($R^{(u)}$) and one for $down$-squarks ($R^{(d)}$). Thus the squark mass
eigenstates ($\tilde{q}_{\alpha }^{\prime }$) and the interaction squark
eigenstates ($\tilde{q}_{\alpha }$) are related by, 
\begin{equation}
\tilde{q}_{\alpha }^{\prime }=\sum R_{\alpha \beta }^{(q)}
\tilde{q}_{\beta}\,,
\label{squarkbs} 
\end{equation}
where explicitly the matrices reads 
\begin{equation}
\tilde{u}_{\alpha }^{\prime }=\left( 
\begin{array}{c}
\tilde{c}_{L} \\ 
\tilde{c}_{R} \\ 
\tilde{t}_{L} \\ 
\tilde{t}_{R}
\end{array}
\right) \,\,,\,\,\tilde{d}_{\alpha }^{\prime }=\left( 
\begin{array}{c}
\tilde{s}_{L} \\ 
\tilde{s}_{R} \\ 
\tilde{b}_{L} \\ 
\tilde{b}_{R}
\end{array}
\right) ,\tilde{u}_{\beta }=\left( 
\begin{array}{c}
\tilde{u}_{1L} \\ 
\tilde{u}_{2L} \\ 
\tilde{u}_{1R} \\ 
\tilde{u}_{2R}
\end{array}
\right) \,\,,\,\,\tilde{d}_{\beta }=\left( 
\begin{array}{c}
\tilde{d}_{1L} \\ 
\tilde{d}_{2L} \\ 
\tilde{d}_{1R} \\ 
\tilde{d}_{2R}
\end{array}
\right) . 
\end{equation}

One obtains the squark mass eigenvalues and eigenstates after the
diagonalization procedure as indicated in Ref.~\cite{herrero}.

\item[b-)] {$u$ and $d$-squarks}
\end{enumerate}

In the symmetric case under $\mathcal{Z}_{2}^{\prime }$ the mass matrix is
trivially diagonal and $\tilde{q}_{3L}$ does not mix with $\tilde{q}_{3R}$,
as a consequence the contribution of the squark sector to the EDM is null.

The interesting case comes from the soft breaking terms of Eq.(\ref{lsoftimpar}). 
For the third generation these terms are given by 
\begin{equation}
M_{Q,3}^{2}\tilde{u}_{3L}^{\star }\tilde{u}_{3L}+M_{u,u}^{2}
\tilde{u}_{3R}^{\star }\tilde{u}_{3R}+A_{33}^{u}\tilde{u}_{3L}
\tilde{u}_{3R}v_{2}+M_{Q,3}^{2}\tilde{d}_{3L}^{\star }\tilde{d}_{3L}+
M_{d,d}^{2}\tilde{d}_{3R}^{\star }\tilde{d}_{3R}+A_{33}^{d}\tilde{d}_{3L}
\tilde{d}_{3R}v_{1}\,\,
\end{equation}
they give the mixing between left-right part of the u-squark and d-squark
sector. This mixing has two important consequences:

The first one is the mass of the squarks of the third family. The off
diagonal entries are proportional to the mass of quarks as shown below, 
\begin{equation}
\mathcal{M}_{\tilde{q}}^{2}=\left( 
\begin{array}{cc}
m_{\tilde{q}_{L}}^{2} & a_{q}m_{q} \\ 
a_{q}m_{q} & m_{\tilde{q}_{R}}^{2}
\end{array}
\right) \;=\;(\mathcal{R}^{\tilde{q}})\left( 
\begin{array}{cc}
m_{\tilde{q}_{1}}^{2} & 0 \\ 
0 & m_{\tilde{q}_{2}}^{2}
\end{array}
\right) \mathcal{R}^{\tilde{q}},  \label{eq:msqmat}
\end{equation}
where $\tilde{q}=\tilde{u},\tilde{d}$. The weak eigenstates $\tilde{q}_{L}$
and $\tilde{q}_{R}$ are thus related to their mass eigenstates $\tilde{q}
_{1} $ and $\tilde{q}_{2}$ by 
\begin{equation}
\left( 
\begin{array}{c}
\tilde{q}_{1} \\ 
\tilde{q}_{2}
\end{array}
\right) =\mathcal{R}^{\tilde{q}}\,\left( 
\begin{array}{c}
\tilde{q}_{3L} \\ 
\tilde{q}_{3R}
\end{array}
\right) ,\hspace{8mm}\mathcal{R}^{\tilde{q}}=\left( 
\begin{array}{rr}
\cos \theta _{\tilde{q}} & \sin \theta _{\tilde{q}} \\ 
-\sin \theta _{\tilde{q}} & \cos \theta _{\tilde{q}}
\end{array}
\right) ,  \label{eq:Rsq}
\end{equation}
with $\theta _{\tilde{q}}$ the squark mixing angle. The mass eigenvalues are
given by 
\begin{equation}
m_{\tilde{q}_{1,2}}^{2}=\frac{1}{2}\left( m_{\tilde{q}_{L}}^{2}+
m_{\tilde{q}_{R}}^{2}\mp \sqrt{(m_{\tilde{q}_{L}}^{2}-m_{\tilde{q}_{R}}^{2})^{2}+
4\,a_{q}^{2}m_{q}^{2}}\,\right) .  \label{eq:sqmasseigenvalues}
\end{equation}
By convention, we choose $\tilde{q}_{1}$ to be the lightest mass eigenstate.
Note that $m_{\tilde{q}_{1}}\leq m_{\tilde{q}_{L,R}}\leq m_{\tilde{q}_{2}}$.
For the mixing angle $\theta _{\tilde{q}}$ we require $0\leq \theta _{\tilde{q}}<\pi $. Thus, we have 
\begin{equation}
\cos \theta _{\tilde{q}}=\frac{-a_{q}m_{q}}{\sqrt{(m_{\tilde{q}_{L}}^{2}-
m_{\tilde{q}_{1}}^{2})^{2}+a_{q}^{2}m_{q}^{2}}},\hspace{8mm}
\sin \theta _{\tilde{q}}=\frac{m_{\tilde{q}_{L}}^{2}-
m_{\tilde{q}_{1}}^{2}}{\sqrt{(m_{\tilde{q}_{L}}^{2}-
m_{\tilde{q}_{1}}^{2})^{2}+a_{q}^{2}m_{q}^{2}}}.
\end{equation}

This mixing is important because it generates contributions to the EDM.

\subsubsection{The masses of Selectrons}

\label{apend:sleptons}

The procedure is the same as in the case of squarks. The mixing in the
selectron sector comes from the following $\mathcal{Z}_{2}^{\prime }$-odd
terms, 
\begin{equation}
M_{L,3}^{2}\tilde{l}_{3L}^{\star }\tilde{l}_{3L}+M_{l,l}^{2}
\tilde{l}_{3R}^{\star }\tilde{l}_{3R}+A_{33}^{l}\tilde{l}_{3L}\tilde{l}_{3R}v_{1},
\end{equation}
The relations among mass eigenstates and interaction eigenstates of
selectron are, 
\begin{equation}
\left( 
\begin{array}{c}
\tilde{e}_{1} \\ 
\tilde{e}_{2}
\end{array}
\right) =\mathcal{R}^{\tilde{e}}\,\left( 
\begin{array}{c}
\tilde{l}_{3L} \\ 
\tilde{l}_{3R}
\end{array}
\right) ,\hspace{8mm}\mathcal{R}^{\tilde{e}}=\left( 
\begin{array}{rr}
\cos \theta _{\tilde{e}} & \sin \theta _{\tilde{e}} \\ 
-\sin \theta _{\tilde{e}} & \cos \theta _{\tilde{e}}
\end{array}
\right) ,  \label{eq:Rsl}
\end{equation}
with $\theta _{\tilde{e}}$ the selectron mixing angle. The mass eigenvalues
are the same as in the case of third family of squarks, therefore their
masses are given by the Eq.(\ref{eq:sqmasseigenvalues}) but with label $q$
instead of $e$.

\subsection{The masses of Gluinos}

It is well known gluinos are the supersymmetric partners of the gluons.
Therefore gluinos are the color octet fermions in the model. On other
hand, as the $SU(3)_{c}$ group is unbroken gluinos can not mix with any
others particles in the model, then they are already mass eigenstates.

Their mass, from Eq.(\ref{lsoftpar}), can be written as 
\begin{equation}
\mathcal{L}_{\mathrm{mass}}^{\mathrm{gluino}}=\frac{m_{\tilde{g}}}{2}
\bar{\tilde{g}}\tilde{g}
\end{equation}
so that its mass at tree level is $m_{\tilde{g}}=|M_{3}|$, where 
\begin{equation}
\tilde{g}^{a}=\left( 
\begin{array}{c}
- \imath \lambda _{C}^{a} \\ 
\imath \overline{\lambda _{C}^{a}}
\end{array}
\right) \,\ ,\hspace{1cm}a=1,\ldots ,8, 
\end{equation}
is the Majorana four-spinor defining the physical gluinos states.

\section{The mechanism of mass generation}

\label{sec:soft}

Once $\mathcal{Z}_{2}^{\prime }$ symmetry is softly broken the fermions are
allowed to interact with their respective superpartners and gluinos (see
at appendix \ref{apend:fsg}). However, the third family is already
disconnected from other two families and we show that the removal of
alignment assumption only changes the content of strange quark mass.

\subsection{Light Fermion Masses}

The $u$-quark can only interact with $u$-squark (defined at Eq.(\ref{eq:Rsq})). 
However, squarks can couple with gluino and also with bino, the
supersymmetric partner of the gauge boson of $U(1)$. First we want to
compare their contribution to the 1-loop mass diagram which generates mass
to the $u$ quark.

In order to estimate their contribution, it is useful to use the
Supersymmetry Parameter Analysis Convention (SPA). Wchich is based on a consistent set
of conventions and input parameters \cite{spa,cmssm,sps}, given
at appendix \ref{sec:spa}. In all the scenarios is easy to see that 
\begin{equation}
g_{s}^{2}m_{\tilde{g}}\gg g^{\prime 2}m^{\prime },
\end{equation}
keep this in mind one can neglect the contribution of the bino.

The interaction between the squarks-gluino-quarks is given by Eq.(\ref{lightquarks}). 
In Fig.(\ref{mass1}) we depict the loop diagram contribution for the
mass of $u$ - quark which gives rise to the following expression as a
function of loop integrals \footnote{$m_{\tilde{u}}$ and $m_{\tilde{g}}$ are
the masses of the u-squark and gluinos respectively.} (see at Eq.(\ref{arhrib})):

\begin{equation}
M_{u}=g_{s}^{2}m_{\tilde{g}}\sin (2\theta _{\tilde{u}}) 
\sum_{\imath =1}^{2}B_{0}(m_{\tilde{u}_{\imath}},m_{\tilde{g}}).  \label{umass1}
\end{equation}

Analogously we obtain for the mass of $d$-quark,  see Fig.(\ref{mass2}) ,the following expression 
\footnote{$m_{\tilde{d}}$ is the d-squark mass.} 
\begin{equation}
M_{d}= g^{2}_{s}m_{\tilde{g}} \sin(2 \theta_{\tilde{d}}) \sum_{\imath =1}^{2}
B_{0}(m_{\tilde{d}_{\imath}},m_{\tilde{g}}).  \label{dmass1}
\end{equation}
These expressions (\ref{umass1},\ref{dmass1}) agree with the results
presented in Refs.\cite{banks,ma,ferrandis1,ferrandis2,cmmc}. 

Likewise the quark case, selectron (Eq.(\ref{eq:Rsl})) interacts with electron
(Eq.(\ref{lltn})) and this interaction is the source for the leading
contribution depicted in Fig.(\ref{mass3}). We obtain the following expression for the
electron mass 
\begin{equation}
M_{e}=g^{\prime 2}\sin (2\theta _{\tilde{e}})m^{\prime}\sum_{\imath =1}^{2}
B_{0}(m_{\tilde{e}_{\imath}},m^{\prime }).  \label{emass1}
\end{equation}

Same as one finds in our first work, the light fermions can couple only with their respective
supersymmetric partners. In contrast, now the strange quark can couple with $s$
and $b$-squark, defined at Eq.(\ref{squarkbs}). This is a source of flavor
non-diagonal sfermion mass matrix.

We define the dimensionless flavor-changing parameters $(\delta^{u,d}_{\imath \jmath})_{AB}$ 
$(A,B = L,R)$ from the flavor off-diagonal
elements of the squark mass matrices ( Eq.(\ref{eq:squarkmassmariana})), in
the following way: first, we set all diagonal entries $M^2_{Q,q}$ and 
$M^2_{u(d),q}$ to be equal to the common value $M^2_{\mathrm{SUSY}}$, then we
normalize the off-diagonal elements to $M^2_{\mathrm{SUSY}}$ \cite{9604378,Chung:2003fi,mariana5,herrero}, 
\begin{eqnarray}
(\delta^{d(u)}_{\imath \jmath})_{AB} = \frac{(M^2_{\tilde{U}(\tilde{D})})_{AB}^{\imath \jmath}}{M^2_{\mathrm{SUSY}}}, 
\;\; (\imath \ne \jmath,\;\; \imath ,\jmath=1,2\;\;\;A,B=L,R).
\label{deltadefb}
\end{eqnarray}

Due this fact the leading contribution to the mass of $s$-quark is shown in
Fig.(\ref{mass4}). Taking into account Eq.(\ref{eq.gluinosqq}) one obtains the
following expression 
\begin{eqnarray}
M_{s} &=&2g_{s}^{2}m_{\tilde{g}}\sum_{\alpha =1}^{2}\left[ 
R_{1\alpha}^{(d)}R_{2\alpha }^{(d)}
B_{0}(m_{\tilde{d}_{\alpha }},m_{\tilde{g}})+R_{1\alpha +2}^{(d)}R_{2\alpha +2}^{(d)}
B_{0}(m_{\tilde{d}_{\alpha +2}},m_{\tilde{g}})\right.   \nonumber  \label{smass1} \\
&+&\left. R_{1\alpha }^{(d)}R_{2\alpha +2}^{(d)}\left( 
\delta _{\alpha \alpha +2}^{d}\right) _{LR}M_{SUSY}^{2}I(m_{\tilde{d}_{\alpha }},
m_{\tilde{d}_{\alpha +2}},m_{\tilde{g}})\right] .  \label{smass1}
\end{eqnarray}
It worth noting that the content of mass of $s$ quark is very different from
the content of others two light quarks, Eqs.(\ref{dmass1},\ref{smass1}), and we are able to make the strange
quark heavier than non-strange quarks.

In fact, even if we consider all the squarks are degenerate in mass the strange quark still is heavier
than non-strange quarks: 
\begin{equation}
m_{s}>4m_{d},
\end{equation}
This relation is in agreement with recent experimental data \cite{pdg} 
\begin{equation}
17\leq \frac{m_{s}}{m_{d}}\geq 22.
\end{equation}

\subsection{Final expressions}

\label{sec:numerical}

As it is clear in Eqs.(\ref{umass1},\ref{dmass1},\ref{emass1}) we have to
perform only one integral. From Eq.(\ref{eqn7034}), we can rewrite the light
fermion mass expressions as follow:
\begin{eqnarray}
M_{u} &=&\frac{g_{s}^{2}m_{\tilde{g}}\sin (2\theta _{\tilde{u}})}{16\pi ^{4}}
\sum_{\imath =1}^{2}\frac{m_{\tilde{u}_{\imath}}^{2}}{(m_{\tilde{u}_{\imath}}^{2}-
m_{\tilde{g}}^{2})}\ln \left( \frac{m_{\tilde{u}_{\imath}}^{2}}{m_{\tilde{g}}^{2}}\right) , 
\nonumber \\
M_{d} &=&\frac{g_{s}^{2}m_{\tilde{g}}\sin (2\theta _{\tilde{d}})}{16\pi ^{4}}
\sum_{\imath =1}^{2}\frac{m_{\tilde{d}_{\imath}}^{2}}{(m_{\tilde{d}_{\imath}}^{2}-
m_{\tilde{g}}^{2})}\ln \left( \frac{m_{\tilde{d}_{\imath}}^{2}}{m_{\tilde{g}}^{2}}\right) , 
\nonumber \\
M_{e} &=&\frac{g^{\prime 2}m^{\prime }\sin (2\theta _{\tilde{e}})}{16\pi ^{4}}
\sum_{\imath =1}^{2}\frac{m_{\tilde{e}_{\imath}}^{2}}{(m_{\tilde{e}_{\imath}}^{2}-
m^{\prime 2})}\ln \left( \frac{m_{\tilde{e}_{\imath}}^{2}}{m^{\prime 2}}\right) .
\end{eqnarray}
These results agree with literature \cite{ma,cmmc,mexico}.

From the scenarios SPA , see Appendix \ref{sec:spa},  the expression 
\begin{equation}
\frac{m_{\tilde{g}}^{2}}{(m_{\tilde{g}}^{2}-m_{\tilde{q}_{\imath}}^{2})}\ln
\left( \frac{m_{\tilde{g}}^{2}}{m_{\tilde{q}_{\imath}}^{2}}\right) 
\label{oba1}
\end{equation}
has  positive values. We can also see that $m_{\tilde{e}_{\imath}}>m^{\prime }$
and therefore we can use the equation above in order to reproduce the mass
pattern of these fermions.

The expression for the mass of $s$ quark has a more complicated integral to
be solved, see Fig.(4) and Eq.(\ref{smass1}) turns into the following: 
\begin{eqnarray}
M_{s} &=&\frac{g_{s}^{2}m_{\tilde{g}}}{16\pi ^{4}}
\sum_{\alpha =1}^{2}\left\{ R_{1\alpha }^{(d)}R_{2\alpha }^{(d)}
\frac{m_{\tilde{g}}^{2}}{(m_{\tilde{g}}^{2}-m_{\tilde{d}_{\alpha }}^{2})}\ln \left( 
\frac{m_{\tilde{g}}^{2}}{m_{\tilde{d}_{\alpha }}^{2}}\right) +R_{1\alpha +2}^{(d)}
R_{2\alpha +2}^{(d)}\frac{m_{\tilde{g}}^{2}}{(m_{\tilde{g}}^{2}-
m_{\tilde{d}_{\alpha +2}}^{2})}\ln \left( \frac{m_{\tilde{g}}^{2}}{m_{\tilde{d}_{\alpha +2}}^{2}}
\right) \right.   \nonumber \\
&+&\left. \frac{R_{1\alpha }^{(d)}R_{2\alpha +2}^{(d)}}{(
m_{\tilde{d}_{\alpha }}^{2}-m_{\tilde{d}_{\alpha +2}}^{2})(m_{\tilde{g}}^{2}-
m_{\tilde{d}_{\alpha }}^{2})(m_{\tilde{d}_{\alpha +2}}^{2}-m_{\tilde{g}}^{2})}\left(
\delta _{\alpha \alpha +2}^{d}\right) _{LR}M_{SUSY}^{2}\left[ 
m_{\tilde{d}_{\alpha }}^{2}m_{\tilde{d}_{\alpha +2}}^{2}\ln \left( 
\frac{m_{\tilde{d}_{\alpha }}^{2}}{m_{\tilde{d}_{\alpha +2}}^{2}}\right) \right. \right.  
\nonumber \\
&+&\left. \left. m_{\tilde{d}_{\alpha }}^{2}m_{\tilde{g}}^{2}\ln \left( 
\frac{m_{\tilde{g}}^{2}}{m_{\tilde{d}_{\alpha }}^{2}}\right) +
m_{\tilde{d}_{\alpha +2}}^{2}m_{\tilde{g}}^{2}\ln \left( 
\frac{m_{\tilde{d}_{\alpha +2}}^{2}}{m_{\tilde{g}}^{2}}\right) \right] \right\} .
\end{eqnarray}
It is important to emphasize that the first two contribution to the mass of
this quark are the same as those in the mass expressions of $u$ and $d$
quarks. The third contribution came from the flavor non-diagonal sfermion
mass matrix contribution. As a result of small mixing, the mass eigenstates are
approximately equal to the flavor eigenstates and hence approximate flavor
eigenstates are propagating in the loop (squarks $\tilde{d}_{1}$ and 
$\tilde{d}_{3}$ or $\tilde{d}_{2}$ and $\tilde{d}_{4}$), this mixing couples squark of
different flavors ($\delta _{13}^{d}$ and $\delta _{24}^{d}$).

\section{\textbf{Conclusions}}

\label{sec:concl}

We showed that the extension of $\mathcal{Z}_{2}^{\prime }$ symmetry to the
squarks sector provide us with a natural mechanism for explaining the chiral
mass hierarchy pattern and also the mass gap between strange and non-strange
quarks. The FCNC problems are under control under $R$-parity invariance
requirements and the breaking of $\mathcal{Z}_{2}^{\prime }$ symmetry only
by SUSY soft terms. There is no need of further assumptions as the alignment
between quark and squark sectors or setting null entries for a particular
mass matrix elements of squarks. The requirement of non-invariance under 
$\mathcal{Z}_{2}^{\prime }$ symmetry for the third family of quarks (squarks)
disconnects this family from the other two families of quarks (squarks). In
the quark sector this disconnection gives rise to the Chiral symmetry
breaking only in the heavy quarks sector ($c$, $t$ and $b$) while the light
quarks remain massless. For squarks sector the family disconnection gives
rise to a particular texture for the mass matrix consistent with
experimental bonds. Once $\mathcal{Z}_{2}^{\prime }$ is softly broken the
light fermions can interact with sfermions and gauginos and they acquire
masses by means of radiative mechanism. Thus we can give a reasonable
explanation of the mass gap between $s$ quark and non strange quarks, even
in the case of all squarks are degenerate in mass 
\footnote{We get naturally $m_{s}>4\times m_{d}$.} at low energy. It is due to the
fact that the $s$ quark can couple with two families of squarks while the $u$
and $d$ quarks can couple only with one family.

\acknowledgments M.C. Rodriguez is supported by Conselho Nacional de Ci\^{e}
ncia e Tecnologia (CNPq) under the contract number 309564/2006-9. C.M.
Maekawa was partially supported by FAPERGS PROADE-2 under contract number
02/1266-6.

\appendix

\section{ Notation}

\label{append:mssm}

In this first appendix we show our notation to the Minimal Supersymmetric
Model (MSSM).

\subsection{The fields of MSSM}

\label{append:not}

\begin{table}[t]
\center
\renewcommand{\arraystretch}{1.5} 
\begin{tabular}{|l|cc|cc|}
\hline
Superfield & Usual Particle & Spin & Superpartner & Spin \\ \hline\hline
\quad$\hat{V}^{\prime}$ (U(1)) & $V_{m}$ & 1 & $\lambda_{B}\,\,$ & 
$\frac {1}{2}$ \\ 
\quad$\hat{V}^{\imath}$ (SU(2)) & $V^{\imath}_{m}$ & 1 & $\lambda^{\imath}_{A}$ & 
$\frac {1}{2}$ \\ 
\quad$\hat{V}^{a}_{c} (SU(3))$ & $G^{a}_{m}$ & 1 & $\tilde{ g}^{a}$ & 
$\frac{1}{2}$ \\ \hline
\quad$\hat{Q}_{\imath}\sim(\mathbf{3},\mathbf{2},1/3)$ & $(u_{\imath},\,d_{\imath})_{L}$ & 
$\frac {1}{2}$ & $(\tilde{ u}_{\imath L},\,\tilde{ d}_{\imath L})$ & 0 \\ 
\quad$\hat{u}^{c}_{\imath}\sim(\mathbf{3^{\ast}},\mathbf{1},-4/3)$ & 
$\bar{u}^{c}_{\imath L}$ & $\frac{1}{2}$ & $\tilde{ u}^{c}_{\imath L}$ & 0 \\ 
\quad$\hat{d}^{c}_{\imath}\sim(\mathbf{3^{\ast}},\mathbf{1},2/3))$ & 
$\bar{d}^{c}_{\imath L}$ & $\frac{1}{2}$ & $\tilde{ d}^{c}_{\imath L}$ & 0 \\ \hline
\quad$\hat{L}_{\imath}\sim(\mathbf{1},\mathbf{2},-1)$ & $(\nu_{\imath},\,l_{\imath})_{L}$ & 
$\frac {1}{2}$ & $(\tilde{ \nu}_{\imath L},\,\tilde{ l}_{\imath L})$ & 0 \\ 
\quad$\hat{l}^{c}_{\imath}\sim(\mathbf{1},\mathbf{1},2)$ & $\bar{l}^{c}_{\imath L}$ & 
$\frac{1}{2}$ & $\tilde{ l}^{c}_{\imath L}$ & 0 \\ \hline
\quad$\hat{H}_{1}\sim(\mathbf{1},\mathbf{2},-1)$ & $(H_{1}^{0},\, H_{1}^{-})$
& 0 & $(\tilde{ H}_{1}^{0},\, \tilde{ H}_{1}^{-})$ & $\frac{1}{2}$ \\ 
\quad$\hat{H}_{2}\sim(\mathbf{1},\mathbf{2},1)$ & $(H_{2}^{+},\, H_{2}^{0})$
& 0 & $(\tilde{ H}_{2}^{+},\, \tilde{ H}_{2}^{0})$ & $\frac{1}{2}$ \\ \hline
\end{tabular}
\renewcommand{\arraystretch}{1}
\caption{Particle content of MSSM.}
\label{tab:mssm}
\end{table}
The particle content of the model is given at Tab.(\ref{tab:mssm}). The
families index for fermions are $\imath , \jmath =1,2,3$. The parentheses in the
first column are the transformation properties under the respective
representation of $(SU(3)_C,SU(2)_L,U(1)_Y)$.

\section{Interaction of Fermion-Sfermion-Gauginos}

\label{apend:fsg}

We present the interactions of sfermions with gauginos.

The interaction between Quark-Squarks-Gluino are given by 
\begin{eqnarray}
\mathcal{L}_{q\tilde{q}\tilde{g}} &=&-\imath \sqrt{2}\left[ \tilde{Q}T^{a}
\bar{Q}\overline{\lambda ^{a}}_{C}-\overline{\tilde{Q}}T^{a}Q\lambda
_{C}^{a}+\tilde{u}^{c}T^{a}\overline{u^{c}}\overline{\lambda ^{a}}_{C}-
\overline{\tilde{u}^{c}}T^{a}u^{c}\lambda _{C}^{a}+\tilde{d}^{c}T^{a}
\overline{d^{c}}\overline{\lambda ^{a}}_{C}-\overline{\tilde{d}^{c}}
T^{a}d^{c}\lambda _{C}^{a}\right]  \nonumber \\
&&
\end{eqnarray}
in the basis of mass eigenstates we rewrite it as follow: 
\begin{eqnarray}
\mathcal{L}_{q\tilde{q}\tilde{g}} &=&-\sqrt{2}\sum_{q=u,d}\bar{q}_{\imath}\left[
U_{\jmath \imath}^{q_{L}\ast }W_{\jmath s}^{\tilde{q}}P_{R}-U_{\jmath \imath}^{q_{R}\ast }
W_{\jmath +3s}^{\tilde{q}}P_{L}\right] T^{a}\tilde{g}^{a}\tilde{q}_{s}+h.c.\,\ ,  \nonumber
\\
&&
\end{eqnarray}
where $T^{a}$ are the $SU(3)_{c}$ generators, 
$P_{L,R}\equiv (1\mp \gamma_{5})/2$, $i,j,s=1,2$ are generation indices. In the gluino interaction, the
flavor changing effects from soft broken terms $M_{\tilde{Q}}^{2}$, 
$M_{\tilde{U}}^{2}$ and $A_{u}$ on the observable are introduced through the
matrix $W^{\tilde{q}}$.

To $u$-squark and $d$-squark we can write 
\begin{equation}
\mathcal{L}_{q\tilde{q}\tilde{g}}=-\sqrt{2}\sum_{q=u,d}g_{s}\,T_{\!rs}^{a}
\left[ \bar{q}_{r}\,(\mathcal{R}_{\imath 1}^{\tilde{q}}P_{R}-
\mathcal{R}_{\imath 2}^{\tilde{q}}P_{L})\,\tilde{g}^{a}\,\tilde{q}_{\imath ,s}^{{}}+hc\right] 
\label{lightquarks}
\end{equation}

The previously introduced intergeneration mixing effects in the squark
sector give rise to strong Flavor Changing effects in processes with neutral
currents through the quark-squark-gluino interaction Lagrangian, which can
now be written in the squark mass eigenstates basis as, 
\begin{eqnarray}
\mathcal{L}_{q\tilde{q}\tilde{g}} &=&-\sqrt{2}g_{s}T_{\alpha u}^{a}\left(
R_{1\alpha }^{(u)\ast }\bar{\tilde{g}}^{a}
\tilde{u}_{\alpha }^{\ast}c_{uL}+R_{1\alpha }^{(d)\ast }\bar{\tilde{g}}^{a}
\tilde{d}_{\alpha }^{\ast}s_{uL}-R_{2\alpha }^{(u)\ast }\bar{\tilde{g}}^{a}
\tilde{u}_{\alpha }^{\ast}c_{uR}-R_{2\alpha }^{(d)\ast }\bar{\tilde{g}}^{a}
\tilde{d}_{\alpha }^{\ast}s_{uR}\right.  \nonumber  \label{eq.gluinosqq} \\
&+&\left. R_{3\alpha }^{(u)\ast }\bar{\tilde{g}}^{a}
\tilde{u}_{\alpha}^{\ast }t_{uL}+R_{3\alpha }^{(d)\ast }\bar{\tilde{g}}^{a}
\tilde{d}_{\alpha}^{\ast }b_{uL}-R_{4\alpha }^{(u)\ast }\bar{\tilde{g}}^{a}
\tilde{u}_{\alpha}^{\ast }t_{uR}-R_{4\alpha }^{(d)\ast }\bar{\tilde{g}}^{a}
\tilde{d}_{\alpha}^{\ast }b_{uR}\right) +h.c  \nonumber \\
&&
\end{eqnarray}
with $\alpha =1,2,3,4$. For simplicity, we will omit the color indices from
now on.

On the other hand, the Feynmann rules between Lepton-Slepton-Gaugino is
computed from 
\begin{eqnarray}
\mathcal{L}_{\tilde{\chi}\tilde{l}l} &=&- \imath \sqrt{2}gT^{\imath}( \tilde{L}\bar {L}
\bar{\lambda}^{\imath}_{A}- \overline{\tilde{L}}L \lambda^{\imath}_{A})- 
\frac{\imath g^{\prime}}{\sqrt{2}}(-1) ( \tilde{L}\bar{L}\bar{\lambda}_{B}- \overline{
\tilde{L}}L \lambda_{B})  \nonumber \\
&-&\frac{\imath g^{\prime}}{\sqrt{2}} 2( \tilde{l}^{c}\overline{l^{c}}
\bar{\lambda }_{B}- \overline{\tilde{l}^{c}}l^{c} \lambda_{B}) \,\ ,
\end{eqnarray}
in terms of masses eigenstates we get the following interaction to
Lepton-Slepton-Neutralino 
\begin{eqnarray}  \label{lltn}
\mathcal{L}_{l\tilde{l}\tilde{\chi}^{0}} &=& \left( 
\overline{\tilde{\chi}^{0}} \right)_{l} (G^{e_{L}}_{\imath sl}P_{L}+G^{e_{R}}_{\imath sl}P_{R}) 
\tilde{e}^{\dagger}_{s}e_{\imath} +h.c. \,\ ,  \nonumber \\
\end{eqnarray}
where 
\begin{eqnarray}
G^{e_{L}}_{\imath sl} &=&G^{e_{L}}_{l}W^{\tilde{e}\star}_{\imath s}- \frac{g}{\sqrt {2}
M_{W}\cos\beta}m_{e_{\imath}}Z^{\star}_{l3}W^{\tilde{e}\star}_{(\imath +3)s},  \nonumber
\\
G^{e_{R}}_{\imath sl} &=&G^{e_{R}}_{l}W^{\tilde{e}\star}_{(\imath +3)s}- 
\frac{g}{\sqrt{2}M_{W}\cos\beta}m_{e_{\imath}}Z_{l3}W^{\tilde{e}\star}_{\imath s}.  \nonumber \\
\end{eqnarray}

\section{SPA Convention}

\label{sec:spa}

The Supersymmetry Parameter Analysis project (SPA)  is a comparative study of supersymmetric particle spectra 
calculated for various SUSY scenarios \cite{spa,cmssm,sps}.  The definition of several scenarios are given at \cite{spa,sps}.

\begin{figure}[ptb]
\parbox{14cm}{
\epsfxsize=50mm
\epsfbox{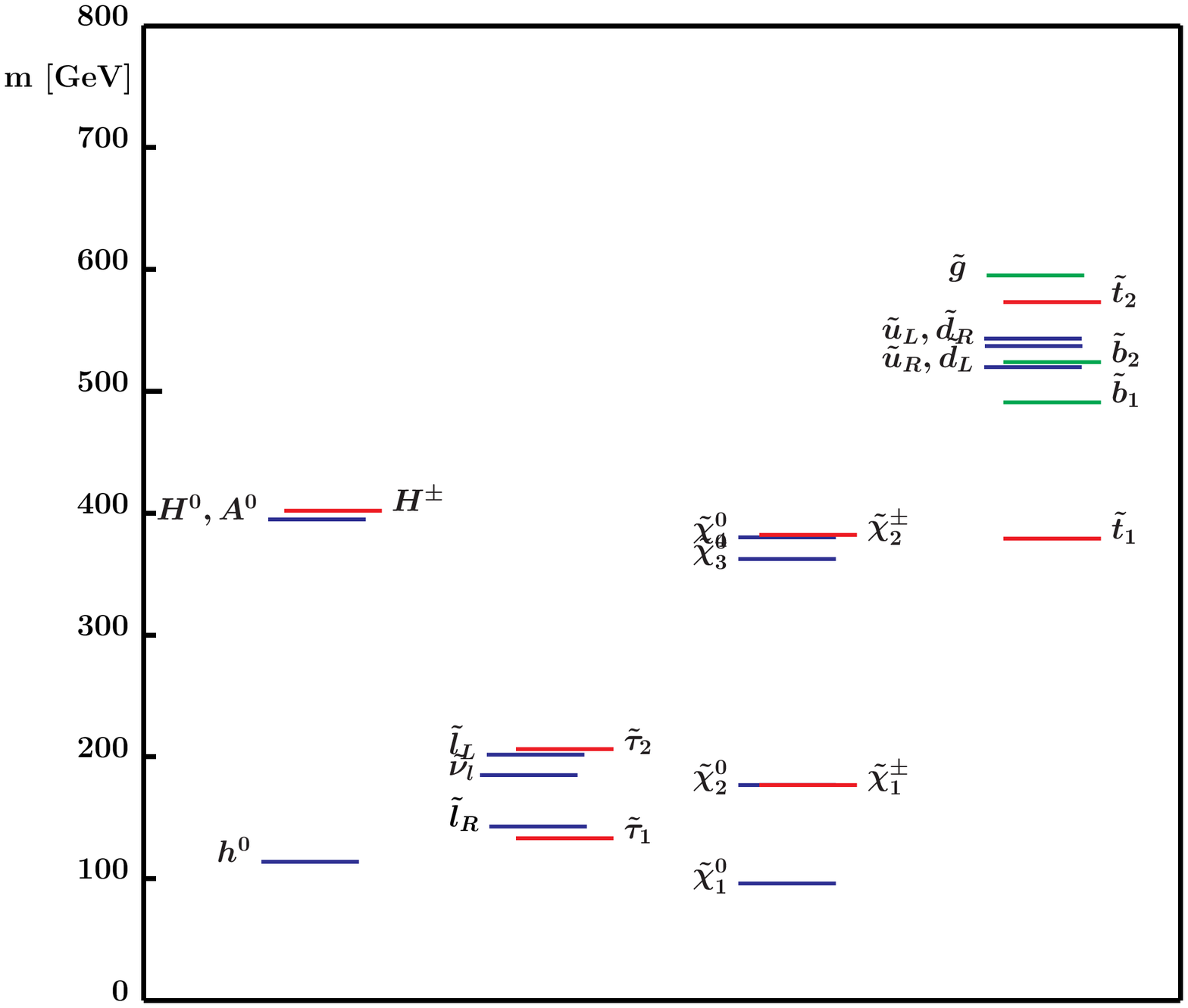}
}\newline
\caption{The SUSY particle spectra for the benchmark points
corresponding to SPS 1a \cite{spa}.}
\label{fig1}
\end{figure}

\begin{figure}[ptb]
\parbox{14cm}{
\epsfxsize=50mm
\epsfbox{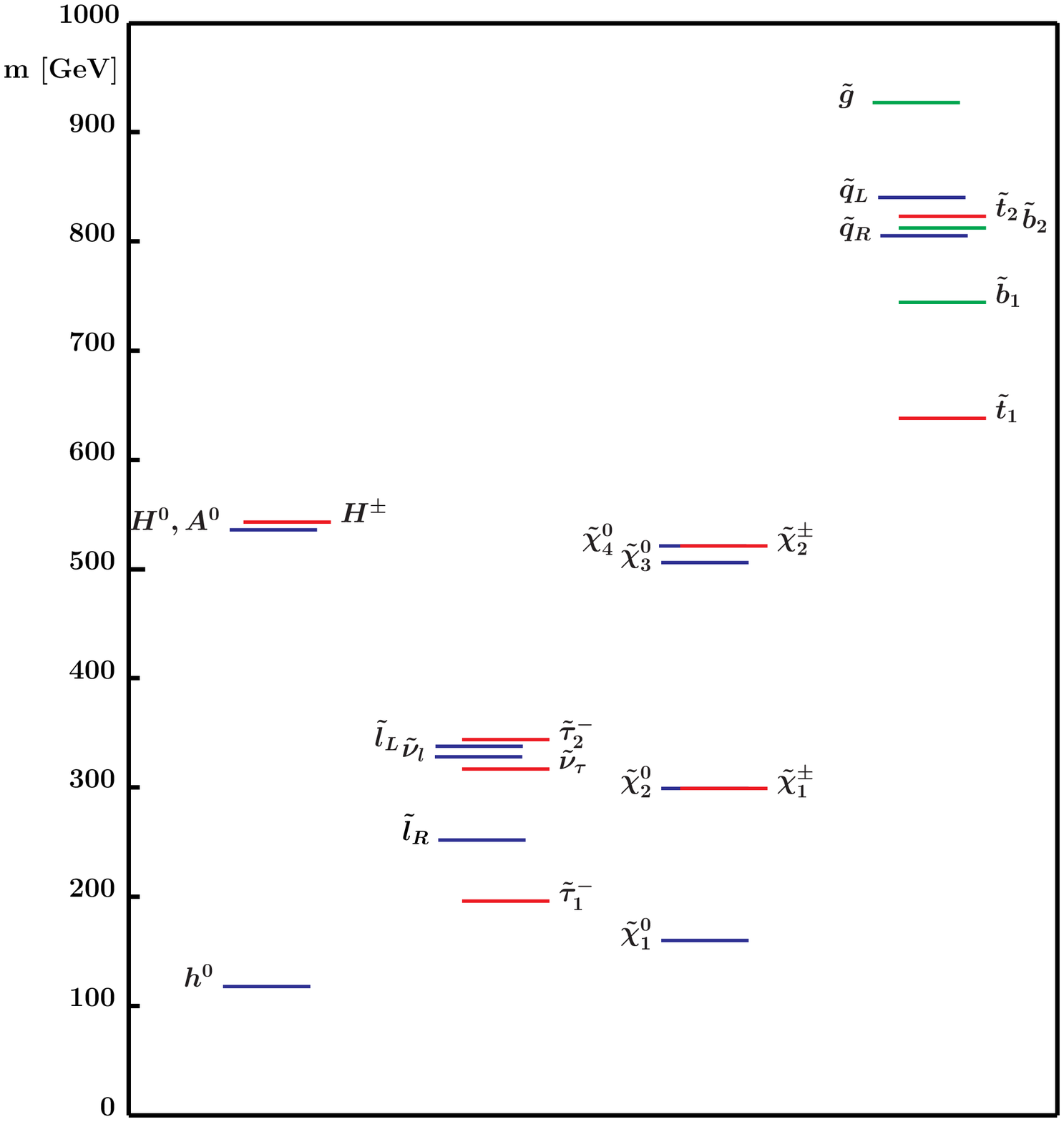}
}\newline
\caption{The SUSY particle spectra for the benchmark points
corresponding to SPS 1b \cite{spa}.}
\label{fig2}
\end{figure}

\begin{figure}[ptb]
\parbox{14cm}{
\epsfxsize=50mm
\epsfbox{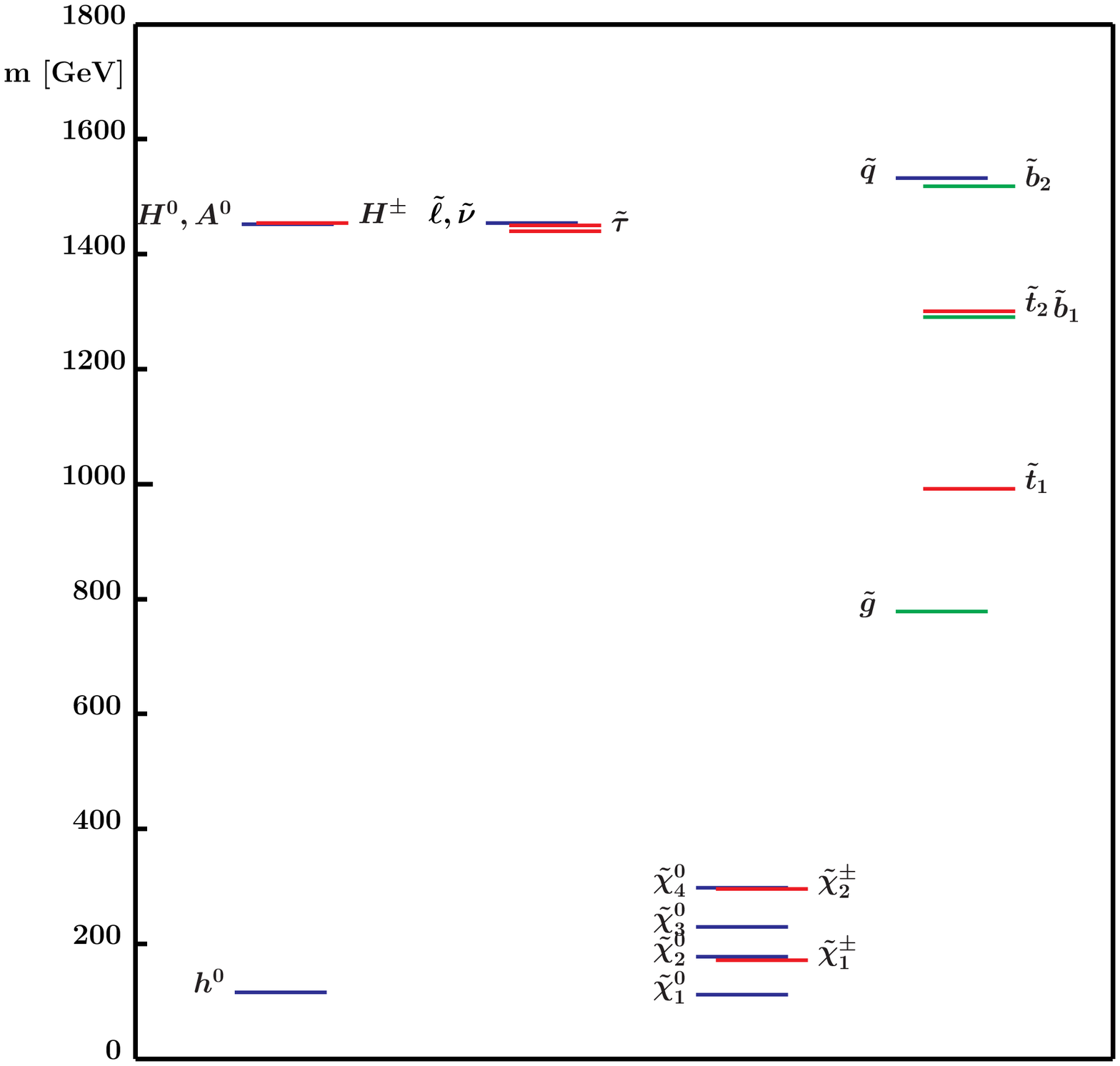}
}\newline
\caption{The SUSY particle spectra for the benchmark points
corresponding to SPS 2 \cite{spa}.}
\label{fig3}
\end{figure}

\begin{figure}[ptb]
\parbox{14cm}{
\epsfxsize=50mm
\epsfbox{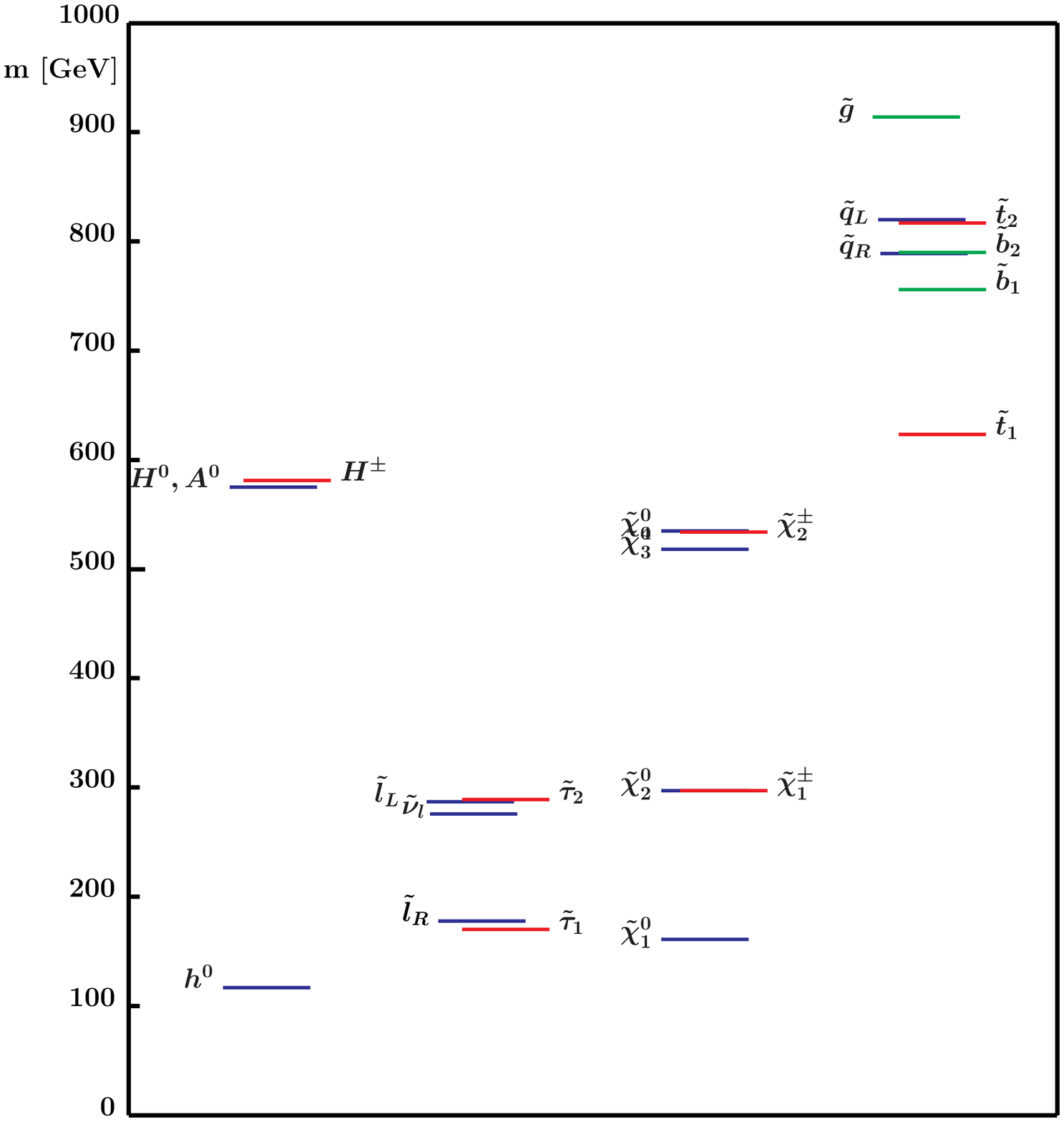}
}\newline
\caption{The SUSY particle spectra for the benchmark points
corresponding to SPS 3 \cite{spa}.}
\label{fig4}
\end{figure}

\begin{figure}[ptb]
\parbox{14cm}{
\epsfxsize=50mm
\epsfbox{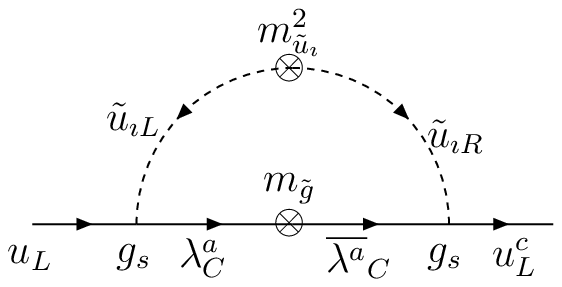}
}\newline
\caption{The diagram which gives mass to quark $u$l, $\lambda^{a}_{C}$ is the gluino while $\tilde {u}
_{i}$, $i=1,2$, is the u-squark.}
\label{mass1}
\end{figure}

\begin{figure}[ptb]
\parbox{14cm}{
\epsfxsize=50mm
\epsfbox{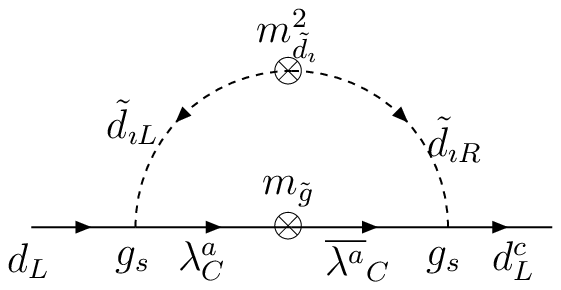}
}\newline
\caption{The diagram which gives mass to quark $d$, $\lambda^{a}_{C}$ is the gluino while $\tilde {d}
_{i}$, $i=1,2$, is the d-squark.}
\label{mass2}
\end{figure}

\begin{figure}[ptb]
\parbox{14cm}{
\epsfxsize=50mm
\epsfbox{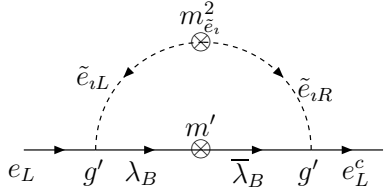}
}\newline
\caption{The diagram which gives mass to electron, $\lambda_{B}$ is the bino while $\tilde {e}
_{i}$, $i=1,2$, is the selectron.}
\label{mass3}
\end{figure}

\begin{figure}[ptb]
\parbox{14cm}{
\epsfxsize=50mm
\epsfbox{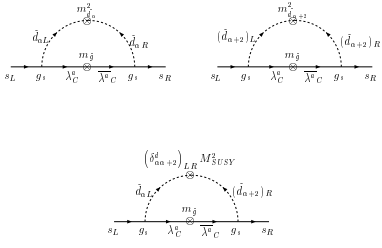}
}\newline
\caption{The diagram which gives mass to quark $s$, $\lambda^{a}_{C}$ is the gluino while $\tilde {s}
_{i}$ and $\tilde{b}_{i}$, $i=1,2$, are the squarks s-squark and sbottom,
respectively.}
\label{mass4}
\end{figure}

The Figs.(\ref{fig1},\ref{fig2},\ref{fig3},\ref{fig4}) show the particle spectra corresponding to SPS1a, SPS1b and SPS3 
\cite{spa}, where the gluinos are the heavy particles. Also in the scenarios SPS5, SPS6, SPS7 and SPS9 the gluinos 
are also the heaviest particles.  Thus, it is simple to show that Eq.(\ref{oba1}) has  positive values. We can also see that $m_{\tilde{e}_{\imath}}>m^{\prime }$
and therefore we can use the equation above in order to reproduce the mass
pattern of these fermions.

At the scenarios SPS2 and SPS8 the gluino are the lightest colored particle and in the last scenario SPS4 we know that 
$m_{\tilde{g}}<m_{\tilde{q}_{L},\tilde{q}_{R}}$ then in both case Eq.(\ref{oba1}) still have 
positive values. 

\section{Feynman integration}

\label{append:feynmanaintegration}

We define the following two point function in following way 
\begin{equation}
B_{0}(p_{1}^{2},m_{1}^{2},m_{2}^{2})=-16\pi \imath \int
 \frac{d^{4}k}{(2\pi)^{4}}\frac{1}{[(k+p_{1})^{2}-m_{1}^{2}][k^{2}-m_{2}^{2}]}\,\ ,
\label{arhrib}
\end{equation}
when the external momentum of the particle is zero, we use the following
convention $B_{0}(0,m_{1}^{2},m_{2}^{2})\equiv B_{0}(m_{1},m_{2})$. Perform
the integral we get the following result \cite{pokorski} 
\begin{equation}
B_{0}(m_{1},m_{2})=1+\ln \left( \frac{{Q}^{2}}{m_{2}^{2}}\right) + 
\frac{m_{1}^{2}}{m_{2}^{2}-m_{1}^{2}}\ln \left( \frac{{m}_{2}^{2}}{m_{1}^{2}}
\right) ,
\end{equation}
where $Q$ is the renormalization scale. After absorb the divergent terms we
can rewrite our result as 
\begin{equation}
B_{0}(m_{1},m_{2})=\frac{m_{1}^{2}}{m_{2}^{2}-m_{1}^{2}}\ln \left(
\frac{{m}_{2}^{2}}{m_{1}^{2}}\right) =\frac{m_{1}^{2}}{m_{1}^{2}-m_{2}^{2}}\ln \left( 
\frac{{m}_{1}^{2}}{m_{2}^{2}}\right) . \label{arhrib1}
\end{equation}
It reproduces the very known results presented at \cite{banks,ma,ferrandis1,ferrandis2,cmmc,mexico,marta}.

Now, we are going to analyze the third integral on Eq.(\ref{smass1}). It is
an integral of the following form 
\begin{eqnarray}
I(m_{1},m_{2},m_{3}) &=& \int\frac{d^{4}p}{(2 \pi)^{4}} 
\frac{1}{(p^{2}-m_{1}^{2})} \frac{1}{(p^{2}-m_{2}^{2})} \frac{1}{(p^{2}-m_{3}^{2})},
\label{util0}
\end{eqnarray}
One uses Eq.(\ref{arhrib}) in order to rewrite Eq.(\ref{util0}) in the following way
\begin{eqnarray}
I(m_{1},m_{2},m_{3})= \frac{1}{m^{2}_{1}-m^{2}_{2}} \left(
B_{0}(m_{1},m_{3})-B_{0}(m_{2},m_{3}) \right).
\end{eqnarray}
We can also apply Eq.(\ref{arhrib1}) to show that 
\begin{eqnarray}
I(m_{1},m_{2},m_{3}) &=& 
\frac{1}{(m^{2}_{1}-m^{2}_{2})(m^{2}_{1}-m^{2}_{3})(m^{2}_{2}-m^{2}_{3})} \left[
m^{2}_{1}m^{2}_{2} \ln \left( \frac{m^{2}_{1}}{m^{2}_{2}} \right) \right. 
\nonumber \\
&+& \left. m^{2}_{1}m^{2}_{3} \ln \left( \frac{m^{2}_{3}}{m^{2}_{1}} \right)
+ m^{2}_{2}m^{2}_{3} \ln \left( \frac{m^{2}_{2}}{m^{2}_{3}} \right) \right].
\label{eqn7034}
\end{eqnarray}
These results are the same as the function $F(x,y,z)$ of Refs. \cite{ferrandis1,ferrandis2}.


\begin{thebibliography}{99}
\bibitem{Sakarov67} A. D. Sakarov, \textsl{JETP Lett.}\textbf{5}, 24 (1967);
I. I. Bigi and A. I. Sanda, \textit{CP Violation}, Cambridge University
Press, United Kindom, (2000).

\bibitem{Weinb79} S. Weinberg, \textsl{Phys. Rev. Lett.}\textbf{21}, 1566
(1979).

\bibitem{exp1} M.Diwan, \textit{et al.}, hep-ex/0608023.

\bibitem{edm} N.F. Ramsey, \textsl{Rep. Prog. Phys.}\textbf{45}, 95 (1982).

\bibitem{israel} A. Gal, \textsl{Phys.Rev.}\textbf{C61}, 028201 (2000).

\bibitem{SM} S. L. Glashow, \textsl{Nucl. Phys.}\textbf{22}, 579 (1961). S.
Weinberg, \textsl{Phys. Rev. Lett}\textbf{19}, 1264 (1967). A. Salam in 
\textit{Elementary Particle Theory: Relativistic Groups and Analyticity},
Nobel Symposium N8 (Alquivist and Wilksells, Stockolm, 1968). S. L. Glashow,
J. Iliopoulos and L. Maini, \textsl{Phys. Rev. }\textbf{D2}, 1285 (1970).

\bibitem{Adelb+85} M. J. Ramsey-Musolf and S. A. Page, hep-ph/0601127. E. G.
Adelberg and W. C. Haxton, \textsl{Ann. Rev. Nucl. Part. Sci,}\textbf{35},
501 (1985), B. Desplanques, J. F. Donohue and B. R. Holstein, \textsl{Ann.
of Phys.}\textbf{124}, 449 (1980).

\bibitem{SMChPT} S.L. Zhu, C. M. Maekawa, B. R. Holstein, M. J.
Ramsey-Musolf and U. van Kolck, \textsl{Nucl. Phys.}\textbf{A748}, 435
(2005), W. H. Hockings and U. Van Kolck, \textsl{Phys. Lett }\textbf{B605},
273 (2005).

\bibitem{ReviewBerez} Z. G. Berezhiani, \textsl{Warsaw Elem.Part.Phys.}, 173
(1993).

\bibitem{Moha+2006} C. Hagedorn, M. Lindner and R. N. Mohapatra, JHEP 
\textbf{0606}, 042 (2006).

\bibitem{dress} M. Dress, R. M. Godbole and P. Royr, \textit{Theory and
Phenomenology of Sparticles} 1st edition, World Scientific Publishing Co.
Pte. Ltd., Singapore, (2004).

\bibitem{tata} H. Baer and X. Tata, \textit{Weak Scale Supersymmetry} 1st
edition, Cambridge University Press, United Kindom, (2006).

\bibitem{9604378} F.~Gabbiani, E.~Gabrielli, A.~Masiero and L.~Silvestrini,
Nucl.\ Phys.\ \textbf{B477}, 321 (1996); M.~Misiak, S.~Pokorski and
J.~Rosiek, Adv.\ Ser.\ Direct.\ High Energy Phys.\ \textbf{15}, 795 (1998);
M.~Ciuchini, E.~Franco, A.~Masiero, L.~Silvestrini, \textsl{Phys.Rev.}
\textbf{D67}, 075016 (2003) [Erratum ibid.\textbf{D68}, 079901 (2003).

\bibitem{Weinberg:1971nd} S.~Weinberg, \textsl{Phys.Rev.}\textbf{D5}, 1962
(1972).

\bibitem{Weinberg:1972ws} S.~Weinberg, \textsl{Phys.Rev.Lett.}\textbf{29},
388 (1972).

\bibitem{Ibanez:1982xg} L.~E.~Ibanez, \textsl{Phys.Lett.}\textbf{B117}, 403
(1982).

\bibitem{banks} T. Banks, \textsl{Nucl. Phys.}\textbf{B303}, 172 (1988).

\bibitem{ma} E. Ma, \textsl{Phys. Rev.}\textbf{D39}, 1922 (1989).

\bibitem{ferrandis1} J. Ferrandis, \textsl{Phys.Rev.}\textbf{D70}, 055002
(2004).

\bibitem{ferrandis2} J. Ferrandis and N. Haba,\textsl{Phys.Rev.}\textbf{D70}
, 055003 (2004).

\bibitem{cmmc} C.M. Maekawa and M. C. Rodriguez, JHEP \textbf{04}, 031
(2006).

\bibitem{Pok+90} H.P. Niles, M. Olechowski and S. Pokorski, \textsl{Phys.
Lett.}\textbf{248}, 378 (1990).

\bibitem{mssm} H. E. Haber and G. L. Kane, \textsl{Phys. Rep.}\textbf{117},
75 (1985).

\bibitem{balin} D. Bailin and A. Love,\textit{Supersymmetric Gauge Field
Theory and String Theory} (Bristol and Philadelphia, Institute of Physics
Publishing, 1999).

\bibitem{mussolf} M.J. Ramsey-Musolf and S. Su, hep-ph/0612057.

\bibitem{Chung:2003fi} D.~J.~H.~Chung, L.~L.~Everett, G.~L.~Kane,
S.~F.~King, J.~D.~Lykken and L.~T.~Wang, \textsl{Phys.Rept.}\textbf{407}, 1
(2005).  [arXiv:hep-ph/0312378].

\bibitem{mariana5} M. Franck and I. Turan, \textsl{Phys.Rev.}\textbf{D74},
073014 (2006).

\bibitem{herrero} A. M. Curiel, M. J. Herrero and D. Temes, \textsl{Phys.Rev.
}\textbf{D67},075008 (2003).

\bibitem{sps} B. C. Allanach \textit{et al}, hep-ph/0202233. See also 
\textbf{http://spa.desy.de/spa}.

\bibitem{spa} J.A. Aguilar-Saavedra \textit{et al}, \textsl{Eur.Phys.J.}
\textbf{C46}, 43 (2006).

\bibitem{cmssm} J.~Ellis, K.~A.~Olive, Y.~Santoso and V.~C.~Spanos, \textsl{
Phys.Rev.}\textbf{D70}, 055005 (2004).

\bibitem{sps} B. C. Allanach \textit{et al}, hep-ph/0202233. See also 
\textbf{http://spa.desy.de/spa}.

\bibitem{pdg} W. -M. Yao \textit{et. al.} (Particle Data Group), \textsl{J.
Phys. G: Nucl. Part. Phys.}\textbf{33}, 1 (2006).

\bibitem{mexico} J.L. Diaz-Cruz, M. Gomez-Bock, R. Noriega-Papaqui and A.
Rosado, hep-ph/0512168.

\bibitem{pokorski} S. Pokorski, \textit{Gauge Field Theories} 2nd edition,
Cambridge University Press, United Kindom, (2000).

\bibitem{marta} S. Davison and M. Losada, JHEP \textbf{0005}, 021 (2000).
\end{thebibliography}
\end{document}